# Unraveling linguistic patterns in dog behaviour

Arunita Banerjee[1], Nandan Das[2], Rajib Dey[2,3], Shouvik Majumder[4], Piuli Shit[1], Ayan Banerjee[3*], Nirmalya Ghosh[3*] and Anindita Bhadra[1*]

[1]Behaviour and Ecology Lab, Department of Biological Sciences

Indian Institute of Science Education and Research Kolkata

Mohanpur Campus, Mohanpur, Nadia

PIN 741235, West Bengal, INDIA

[2]Tissue Optics and Microcirculation Imaging, School of Physics, National University of Ireland, Galway

[3]Department of Physical Sciences

Indian Institute of Science Education and Research Kolkata

Mohanpur Campus, Mohanpur, Nadia

PIN 741235, West Bengal, INDIA

[4]Department of Mathematical Sciences

Indian Institute of Science Education and Research Kolkata

Mohanpur Campus, Mohanpur, Nadia

PIN 741235, West Bengal, INDIA

[*] Corresponding Authors

Apparently random events in nature often reveal hidden patterns when analysed using diverse and robust statistical tools. Power-law distributions, for example, project diverse natural phenomenon, ranging from earthquakes[1] to heartbeat dynamics[2] onto a common platform of statistical self-similarity. A large range of human languages are known to follow a specific regime of power-law distributions, the Zipf-Mandelbrot law, in addition to showing properties like the Pareto principle and Shannon entropy[3,4]. Animal behaviour in specific contexts have been shown to follow power-law distributions[5,6]. However, the entire behavioural repertoire of a species has never been analysed for the existence of underlying patterns. Here we show that the frequency-rank data of randomly sighted behaviours at the population level of free-ranging dogs follow a scale-invariant power-law behaviour. While the data does not display Zipfian trends, it obeys the Pareto principle and Shannon entropy rules akin to languages. Interestingly, the data also exhibits robust self-similarity patterns at different scales which we extract using multifractal detrended fluctuation analysis[7]. The observed multifractal trends suggest that the probability of consecutive occurrence of behaviours of adjacent ranks is much higher than behaviours widely separated in rank. Since we observe such robust trends in random data sets, we hypothesize that the general behavioural repertoire of a species is shaped by a syntax similar to languages. This opens up the prospect of future multifractal and other statistical investigations on true time series of behavioural data to probe the existence of possible long and short-range correlations, and thereby develop predictive models of behaviour.

Simon de Laplace, one of the pioneers of probability theory, believed that events look random to us only because we are limited by our ability to grasp the numerous hidden factors that influence, and thereby affect such events[8]. An interesting commonality that has emerged across studies on many apparently disparate phenomena, from the distributions of species in specific habitats to connections between nodes in social networks, is the presence of a power-law distribution[9]. Power-law distributions represent classic cases of order in chaos and are often considered as a signature for the presence of underlying mechanisms that lead to the observed data[10].

Human languages have been known to demonstrate such probabilistic distributions. Specifically, the frequency of occurrence of the words from a piece of text mostly follow a power-law distribution obeying the Zipf-Mandelbrot law[11]. Analogous to human communication through languages, are animal communication systems exhibited by vocalizations and behavioural displays. However, communication does not necessarily involve vocalizations alone, and can be achieved through more complex and subtle signals like pheromones, postures, gaits, and even a combination of these[12]. Hence, the behavioural repertoire of a species might contain hidden patterns, which carry important information that is not revealed easily to an observer. In fact, power-law distributions have been observed in various behaviours in the context of movement, such as foraging patterns of animals[13,14], free-flight of *Drosophila*[15], swarming behaviour[16], and even human mobility[17]. These behavioural patterns can be approximated by Lévy flights[14, 16], and also display scale-free characteristics.

While the scale-free nature of power-law distributions in the living world is quite ubiquitous, a scale invariant fractal-like nature has also been reported in a few observations of behaviour, such as sleep-wake cycle transitions[18], social behaviour of wild chimpanzees[19], diving patterns in seabird foraging[20], etc. A study on Japanese macaques showed that the behavioural patterns of these primates display a fractal nature, and parasite load affects the level of complexity of their behaviour[21]. The fractal nature of behavioural signatures can thus be used to understand the behavioural patterns of animals, as well as to identify anomalous behaviour in species of interest. However, monofractal measures that are characterized by a single scaling exponent are rarely observed in natural phenomena, which rather exhibit more complex scaling behaviour comprising of a multitude of scaling exponents corresponding to many constituent interwoven fractal subsets. This special class of complex self-affine processes – termed as multifractals[7] – have been observed and studied in a wide variety of natural phenomena, structures and processes, which include the physiological time series of heartbeats[2], the sun's magnetic field dynamics[22], turbulence phenomenon[23], swimming behaviour in zooplanktons[24], and so on. The signatures of multifractality are more robust in cases where its origin is due to a broad probability distribution, so that it cannot be removed by random shuffling of the fluctuation series, as is the case for statistical fractals. Thus, in the case of the former, the quantification of multifractality provides potentially invaluable information on the underlying complex correlations and scaling behaviour – which can enable entirely new perspectives in the analysis of any natural phenomena exhibiting self-similarity[25].

It is therefore evident that the analysis of power-law and scaling relationships in behavioural data can be used to identify the existence of universal principles within the seemingly arbitrary behavioural repertoire of a species. While specific behavioural categories like locomotion, foraging and vocalizations have been subjected to such analysis, to the best of our knowledge, the entire behavioural repertoire of a species has never been tested for linguistic features. Dogs are the first species to have been domesticated by humans, and they share a long history of co-evolution with our species[26]. Free-ranging dogs, which contribute to nearly 80% of the world's dog population, can provide interesting insights into the biology of dogs in general, and the evolution of the dog-human relationship in particular. Many studies attempt to understand the communication between dogs and humans using vocalizations and gestures[27]. Free-ranging dogs are social, with groups displaying interesting cooperation-conflict dynamics[28,29] and individuals showing various degrees of socialization with humans[30]. In India, the free-ranging dogs are ubiquitous, experiencing a wide variety of interactions with humans, from very positive to very negative[31]. They spend a large proportion of time in inactivity[32] and their activity patterns are not easy to predict. Using random, population-level sampling of behaviour, we investigated whether the free-ranging dog behavioural repertoire has any inherent language-like pattern.

We obtained data on 5669, 1308 and 506 random sightings of free-ranging dogs during three different sampling bouts in various parts of India (**Supplementary Table S1**). The number of behaviours observed in these three data sets were 83, 36 and 29 respectively, summing up to a total of 93 unique behaviours in the combined data set. The frequency of observed behaviours plotted against their ranks showed power-law distributions of the nature $P(r) = r^{-\alpha}$, with the log-log plots having slopes of $\alpha \sim 1.73$ (**Table 1; Figure 1a; 1b**). Thus, though the behavioural repertoire showed a general power-law distribution, it did not follow the Zipf-Mandelbrot law, which requires the data to fit a slope of -1 in the log-log scale. Though the data did not seem to follow one of the prime signatures of linguistics, it did show agreement with the Pareto principle[3], with 80% of the cumulative proportion of the time activity budget being explained by 20% of the observed behaviours (on a normalized scale) (**Supplementary Figure S1a**). The Shannon entropy of the behavioural data scaled with the frequency of occurrence of the behaviours (**Supplementaty Figure S1b**), suggesting that the least frequent behaviours provided the most information about the system, which is also true for languages[4].

Since power-laws are typically scale-invariant, we checked for scaling in our data using the three data sets and their all possible combinations (**Supplementary Table S1**). Indeed, our data showed scale-invariance, with the slopes being very similar (range: -1.717 to -1.822), in spite of the widely differing population sizes (**Figure 1c; 1d**). However, the scaling behaviour was rather intriguing as the slope $\alpha$ was not uniform throughout the entire range of rank $r$ (**Figure 2a**), with different values for $\alpha$ over three selected $r$ –ranges: low ($r = 1\ to\ 21$), intermediate ($r = 21\ to\ 56$) and high ($r = 51\ to\ 93$). This is in contrast to the value obtained for $\alpha$ (~1.81) while fitting over the entire range of the ranks for the combined data set. The corresponding representative behavioural fluctuation series $\xi(n)$ numerically generated using a single power-law probability approximation of the frequency vs rank data was then subjected to MFDFA analysis (see Methods). The derived generalized Hurst exponent $h(q)$ exhibited a bi-fractal scaling behaviour $\left[h(q)\sim\frac{1}{q} for (q > (\alpha - 1)) \text{ and } h(q)\sim\frac{1}{(\alpha-1)} for (q \leq (\alpha - 1))\right]$ (see **Supplementary Figure S2**), which is expected for an uncorrelated random fluctuation series generated using a power-law probability distribution[7]. Next, we proceeded to generate a fluctuation series $\xi(n)$ numerically using multiple power-law probability distributions (see Methods) and once again subjected the series to MFDFA analysis. The observed wide range of large and small fluctuations in the detrended fluctuation series underscores the complex nature and the overall randomness of the behavioural fluctuations (**Figure 2b**). The MFDFA-derived moment ($q$) dependence of the fluctuation functions $logF_q(s)$ vs $\log s$ (**Figure 2c**) indicates the presence of multifractal scaling, as the slopes vary significantly for the entire range of $q$. The resulting continuous variations of $h(q)$ and $\tau(q)$ with varying $q$ furnishes concrete evidence of multifractality (**Figure 2d**), with the variations of $h(q)$ being more prominent for negative values of $q$ as compared to positive values. This is in sharp contrast to that observed for the fluctuation series with single power-law probability (see **Supplementary Fig. S1c**). These multifractal trends suggest that consecutive occurrence of behaviours of adjacent ranks (representing small fluctuations that are captured by the negative $q$) dominate the overall scaling behaviour as compared to the behaviours that are widely separated in rank (representing large fluctuations that are captured by the positive $q$). The corresponding strength of multifractality is subsequently quantified via the width ($\Delta\beta$) of the singularity spectrum $f(\beta)$ (**Figure 2e**), with the reasonably large magnitude of $\Delta\beta (= 1.12)$ clearly demonstrating strong multifractality in the behavioural data of free-ranging dogs.

The implication of our findings in the context of the analysis of behavioural data is rather intriguing. The fact that the slope of the power-law fit to the rank-frequency plot of the data is non-Zipfian, while the requirements of the Pareto principle and Shannon entropy are satisfied by the data, is actually not a contradiction in terms of drawing parallels of dog behaviour with languages. This is due to the fact that non-Zipfian behaviour has also been found in a language when larger corpora have been used to account for the large diversities in the word pool that constitute the language[33]. Indeed, we strongly believe that a multi-fractal analysis of languages with word pools drawn from a corpora instead of a single corpus would lead to traits similar to the observations we report here. This also tempts us to conclude that patterns observed in behavioural repertoire may be very similar to that in languages, provided both are analysed in a similar fashion. However, the information obtained from the multifractal analysis of the behavioural repertoire of a species is indeed of a richer hue. In the present context, our results imply that for a single dog, the probability of consecutive occurrence of behaviours of adjacent rank is much higher than behaviours widely separated in rank. This inference is quite remarkable, given that the data pool is drawn from a large number of randomly sighted behaviours at the population level. From a purely probabilistic standpoint, this would demand that the probability of occurrence of a high-frequency behaviour (high rank) would be very large immediately after a low frequency behaviour (low rank). Specifically, a behaviour such as sleeping (Rank 1, highly probable) should follow barking (Rank 18, less probable). However, this is rarely expected to occur in practice, which is also validated by our multifractal analysis. The proposed multifractal analysis in combination with appropriate probabilistic models may in principle be able to *predict* a behaviour at a later time when the behaviour at a certain time is known. To achieve that goal, we intend to extend this study towards the true time series of behavioural data of free-ranging dogs to investigate the existence of possible long and short-range correlations and their effects on the resulting multifractal trends. The resulting short-range correlations can be exploited to predict immediate behavioural sequences, while long-range correlations might provide insights into personalities of individuals. Such information would be of great value in behavioural studies and prove to be most useful in mitigating dog-human conflict. Our method is also completely general in nature, and can, in principle, address similar behavioural questions in any species.

**Figures**

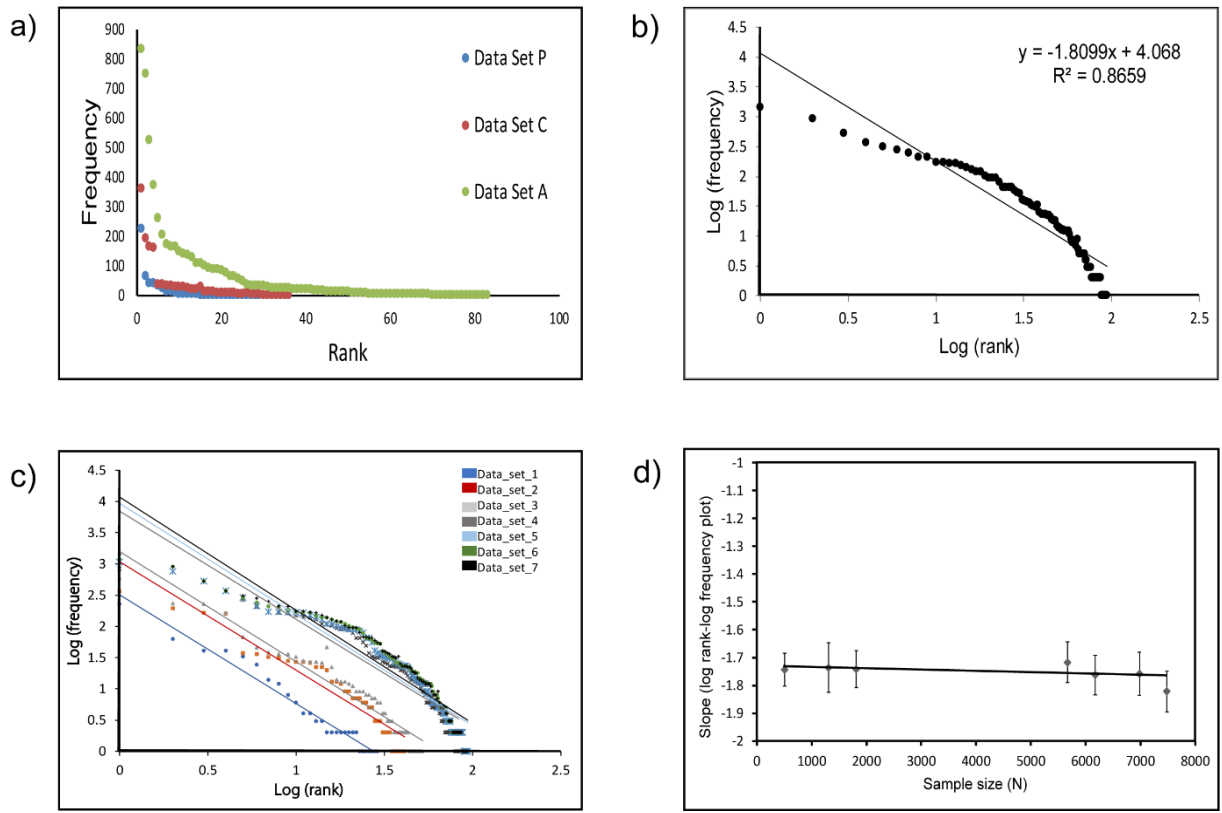

Figure 1: a) scatterplots showing the frequency distributions of observed behaviours arranged according to rank in the three data sets (N = 506, 1308 and 5669 for P, C and A respectively). b) The black dots represent the frequency-rank distribution plotted on a log-log scale for all 7482 observations (data sets P+C+A), which yielded 94 unique ranks. The line represents the linear fit to the curve represented by the equation y = -1.8099x + 4.068, having a $R^2$ value of 0.8659. c) The frequency vs rank plots (on a log-log scale) of the three data sets and all their possible combinations, yielding 7 data sets. Each colour represents a data set and the straight line of the corresponding colour represents the linear fit to the data. d) A scatterplot showing the mean (dots) and standard errors (lines) of the slopes of the log-log plots as shown in "c". The nearly straight line parallel to the x-axis fitting the data suggests scale invariance with change in sample size for this data.

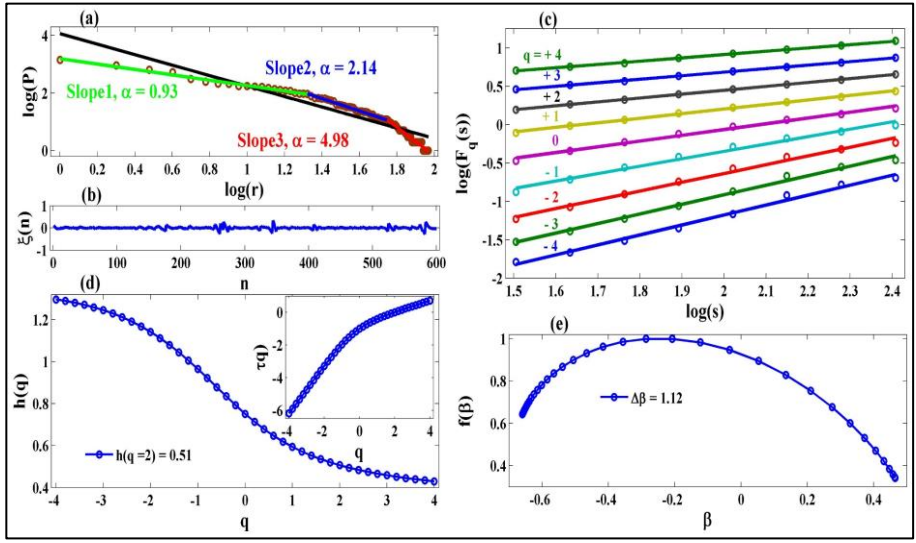

**Figure 2:** *Multifractal patterns in the statistically equivalent behavioural random fluctuation series generated using multiple power-law approximation of the Frequency vs behavioural rank data.* **(a)** Power-law fitting of the Frequency ($P$) vs rank ($r$) data of dog behaviour (shown in natural logarithm scale). Fitting at three selected $r$-ranges (lower (red), intermediate (blue) and higher (green), respectively) yield different values for the scaling exponent (indicating multifractality). Overall fitting over the entire range of the rank with a single power-law ($\alpha = 1.81$) is shown by black line. **(b)-(e):** Results of the MFDFA analysis on the fluctuation series $\xi(n)$ generated using ***multiple power-law approximation*** of frequency vs rank data. **(b)** The detrended (by least square polynomial fitting) fluctuation series representing the behavioural fluctuations $[\xi(n)]$. **(c)** The log-log plot of the moment ($q$ = -4 to +4) dependent fluctuation function $F_q(s)$ vs $s$. Considerable variations in the slopes for the entire range of $q$ values indicate the presence of multifractal scaling. **(d)** The variation of generalized Hurst exponent $h(q)$ (derived from the slopes of $\log F_q(s)$ $vs$ $\log(s)$) and $\tau(q)$ (derived using Eq. 3, shown in inset) with varying $q$. Continuous variations of $h(q)$ and $\tau(q)$ with varying $q$ confirm multifractality. **(e)** The corresponding multifractality is quantified via the singularity spectrum $f(\beta)$ (derived using Eq. 4) and its full width $\Delta\beta$ is noted.

## Methods

**Sampling:**

The data was collected through instantaneous scan sampling of free-ranging dogs in urban and semi-urban habitats. The observer traversed along a pre-decided route, at random time points during the day, on different days. Whenever one or more dogs were sighted, the age class (adult/juvenile), sex, instantaneous behaviour at the time of sighting for each dog were recorded along with the date, time and location of sighting[32]. Three data sets, collected at different times from different locations, were used for this analysis (**Supplementary Table S1; Figure S3**).

**Data Analysis:**

*Linguistic patterns of behaviour*

The frequency (number of occurrences) of each behaviour was estimated from the data sets. For each data set, the behaviours were ranked according to their frequencies of occurrence, with the most frequent behaviour being rank 1. The behavioural frequencies were plotted against their respective ranks, to check for a power law distribution. The frequency and rank were re-plotted on a log-log scale and the slope of the resulting trend line was considered to check for a fit to the Zipf-Mandelbrot law[3,4]. The three data sets and all their possible combinations were used to check for scale invariance.

The behavioural ranks were normalized on a scale of 1-100 for each data set separately, and the cumulative time spent in each behaviour was calculated, such that the total time was 1. The behaviours contributing to the first 80% of time spent were identified for each data set and the distributions were tested for adherence to the Pareto principle. The Shannon entropy was estimated for the data sets and plotted against the ranks of the behaviours to check if the frequency of occurrence influenced the Shannon entropy.

*Generation of the statistically equivalent behavioural random fluctuation series from the frequency vs rank data*

The collected behavioural data of free-ranging dogs were ranked according to their frequencies ($P$) of occurrence, with the most frequent behaviour being ranked $r = 1$. The corresponding Frequency vs rank data (for the range $r = 1\ to\ 93$) was first fitted to a single power-law distribution $P(r) \sim r^{-\alpha}$, yielding a value of $\alpha = 1.81$ (shown in Fig. 2a). A series of random numbers that statistically represent the random sequence of different behavioural events (with rank $r$) were generated through sampling of the random numbers by the resulting power-law distribution. The numerical ranks of the representative behaviours in the series were represented by an appropriately scaled (by the mean and the standard deviations of rank) random variable $\xi(n)$, with $n$ representing any random sequence of behavioural event.

However, the actual frequency vs rank $(P(r))$ behavioural data exhibited three different power-law coefficients $\alpha$ for three different range of ranks (Fig. 2a). The corresponding series $\xi(n)$ was therefore obtained by sampling the random numbers with power-law distributions with three different values for $\alpha$ (0.93, 2.14, 4.98) and then by random shuffling of the generated sequences. This process of generation of the uncorrelated random fluctuation series using multiple power-law probability distribution is specifically applicable here because the behavioural data were not collected in any given sequence of time rather these were collected at random time points. Thus, in this scenario, the $\xi(n)$ series of random events synthesized using the aforementioned method represents the actual random behavioural fluctuation data in a statistical sense. The $\xi(n)$ fluctuation series generated using either a single or multiple power-law probability distributions were subsequently subjected to multifractal detrended fluctuation analysis.

*Multifractal detrended fluctuations analysis*

A statistical monofractal series is one whose variance exhibits a power-law scaling described by a single scaling exponent, namely, Hurst exponent, $H$ ($0<H<1$)[34]. A multifractal series, on the other hand, exhibits complex scaling behaviour comprising of many interwoven fractal subsets characterized by different local Hurst exponents[7]. Multifractal detrended fluctuation analysis (MFDFA) is a generalized approach to characterize such complex multi-affine processes[35]. Using this approach, the fluctuation profile $\xi(n)$ (series of length N, $n = 1 \ldots \ldots N$) is first divided into $N_s$ = int (N/$s$) segments $m$ of equal length $s$. The local trends ($y_m(n)$) of each segment $m$ are determined by polynomial fitting. The fitted trends are then subtracted from the profile to obtain the detrended fluctuations and the corresponding variance of a segment is subsequently obtained as

$$F^2(m,s) = \frac{1}{s}\sum_{n=1}^{s}[Y\{(m-1)s+n\} - y_m(n)]^2 \quad (1)$$

The variances are then averaged over all the segments to construct the moment ($q$) dependent fluctuation function

$$F_q(s) = \left\{\frac{1}{2N_s}\sum_{m=1}^{2N_s}[F^2(m,s)]^{\frac{q}{2}}\right\}^{1/q} \quad (2)$$

In order to quantify the scaling behaviour, the fluctuation function is approximated to follow a power-law scaling $F_q(s) \sim s^{h(q)}$. The multifractality (if any) of the signal is subsequently characterized via the moment dependence of the generalized Hurst scaling exponent $h(q)$, the classical multifractal scaling exponent $\tau(q)$, and the singularity spectrum $f(\beta)$. These are related as

$$\tau(q) = qh(q) - 1 \qquad (3)$$

$$\beta = \frac{d\tau}{dq}, \ f(\beta) = q\beta - \tau(q) \qquad (4)$$

where $\beta$ is the singularity strength and the full width of $f(\beta)$, $\Delta\beta$ (taken at $f(\beta) = 0$) is a quantitative measure of the strength of multifractality. Note that $h(q = 2)$ corresponds to the Hurst exponent $(H)$ of an equivalent monofractal series.

**Author contributions**

ArB collected all data for data set 4, PS collected data for data set 1, data set 2 has been previously published from the Dog Lab, the authors would like to thank Ms. Sreejani Sen Majumder and Dr. Ankita Chatterjee for their participation in this earlier work, for which AB was the corresponding author. ND and RD performed the MFDFA under the supervision of NG. SM participated in the generation of the randomized data for the MFDFA. AyB provided inputs on the data analysis. ArB carried out all the analysis other than the MFDFA. AB conceptualized the problem, supervised the data collection and all analysis other than the MFDFA. AB, ArB, AyB and NG co-wrote the manuscript.

**Funding statement**

ArB was supported by a fellowship from the Indian Institute of Science Education and Research (IISER) Kolkata, PS was supported by a DST INSPIRE fellowship.

**Supporting Information**

| Combinations | N | Renamed |
|---|---|---|
| Data Set P | 506 | Data_set_1 |
| Data Set C | 1308 | Data_set_2 |
| Data Set P+C | 1814 | Data_set_3 |
| Data Set A | 5669 | Data_set_4 |
| Data Set P+A | 6175 | Data_set_5 |
| Data Set A+C | 6977 | Data_set_6 |
| Data Set A+P+C | 7482 | Data_set_7 |

**Table 1:** Table showing the seven combinations prepared from three principal data sets (named P, C and A) of the lab. N denotes the sample size of each data set. The right-most column gives the name of the data set as used in the analysis. Data set C has been used earlier in the paper Sen Majumder et al (2014)[32].

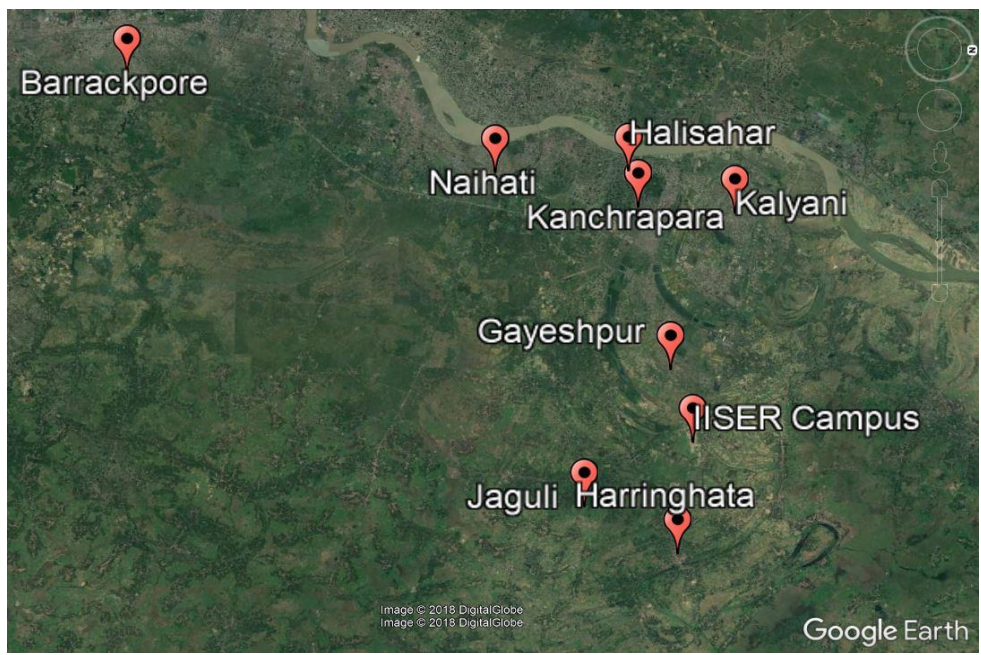

**Figure S1:** Part of a map prepared using Google Earth©, showing locations in which sampling for the study was carried out.

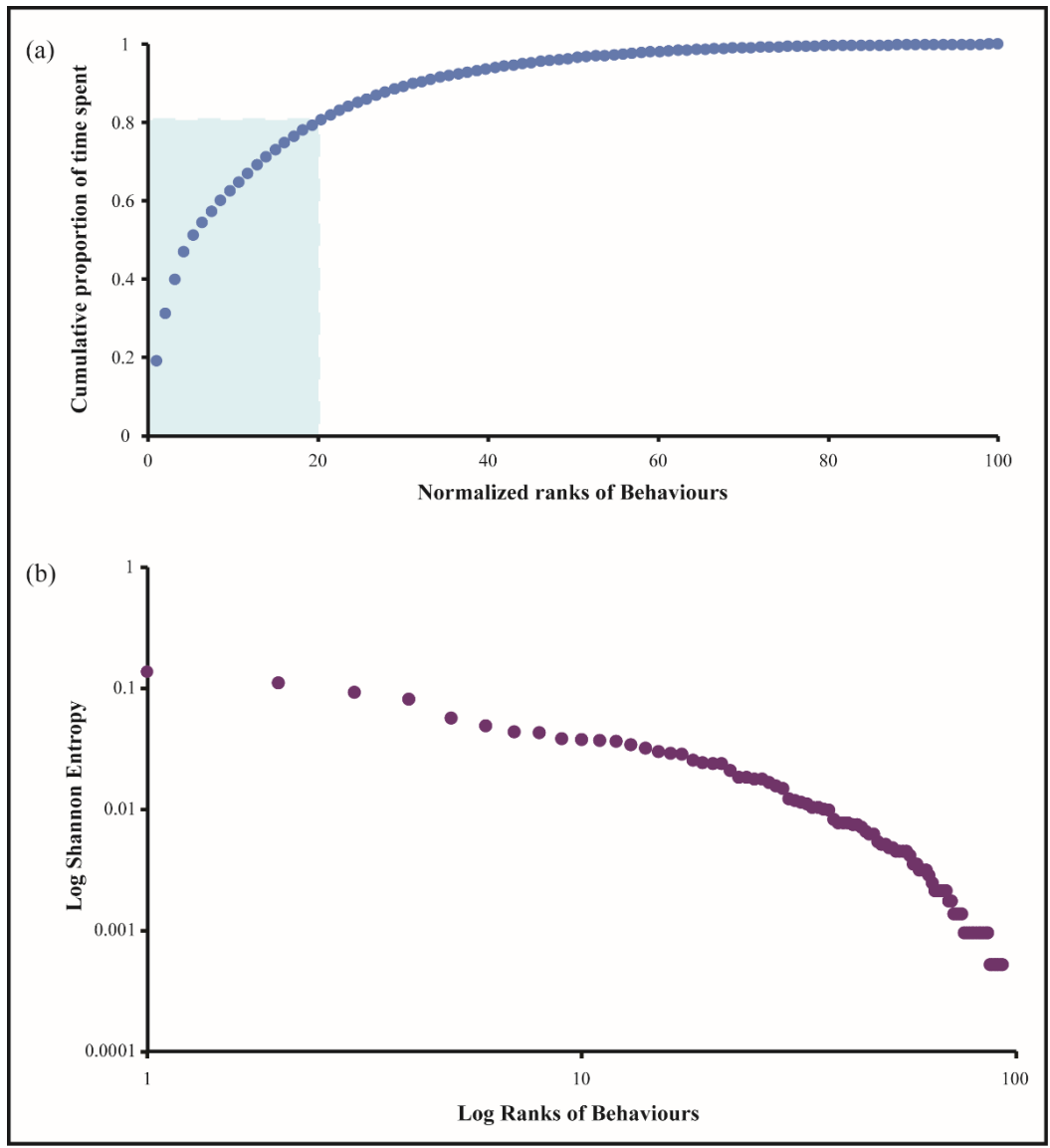

**Figure S2:** a) A scatterplot showing the cumulative proportion of time spent in behaviours (time-activity budget) against the normalized ranks of behaviours, for the pooled data (N = 7482). The highlighted area in pale blue shows the concurrence with the 80-20 principle or Pareto principle. b) A scatterplot showing the Shannon entropy for behaviours arranged according to their ranks in a log-log scale.

**Supplementary text**

As discussed in the main text, the behavioural fluctuation series $\xi(n)$ was first generated using a single power-law probability approximation (with $\alpha \sim 1.81$) of the frequency vs rank data. This was subjected to the MFDFA analysis. The corresponding results are summarized in **Figure S3**. Variations in the slopes of $\log F_q(s)$ vs $\log s$ (**Figure S3a**) with varying moment $q$ provide evidence of multitude of scaling. However, variations in the slopes are observed to be more prominent for positive $q$-values as compared to negative $q$-values. Accordingly, the derived generalized Hurst exponent $h(q)$ exhibit gradual variations with $q$ for $q > (\alpha - 1)$ and nearly uniform values of $h(q)$ for $q \leq (\alpha - 1)$ (**Figure S3b**). These variations of $h(q)$ (and the corresponding variations of the classical scaling exponent $\tau(q)$, shown in inset) are characteristic features of bi-fractal scaling behaviour[35]. This bi-fractal scaling behaviour was further verified on fluctuation series $\xi(n)$ generated using single power-law probability distributions for different other values of $\alpha$ also. These results of MFDFA analysis and that with multiple power-law probability distributions (Figure 2 of main text) confirmed that the multifractal trends in the behavioural data are exclusively due to the presence of multiple power-law scaling exponents $\alpha$ for different ranges of the rank.

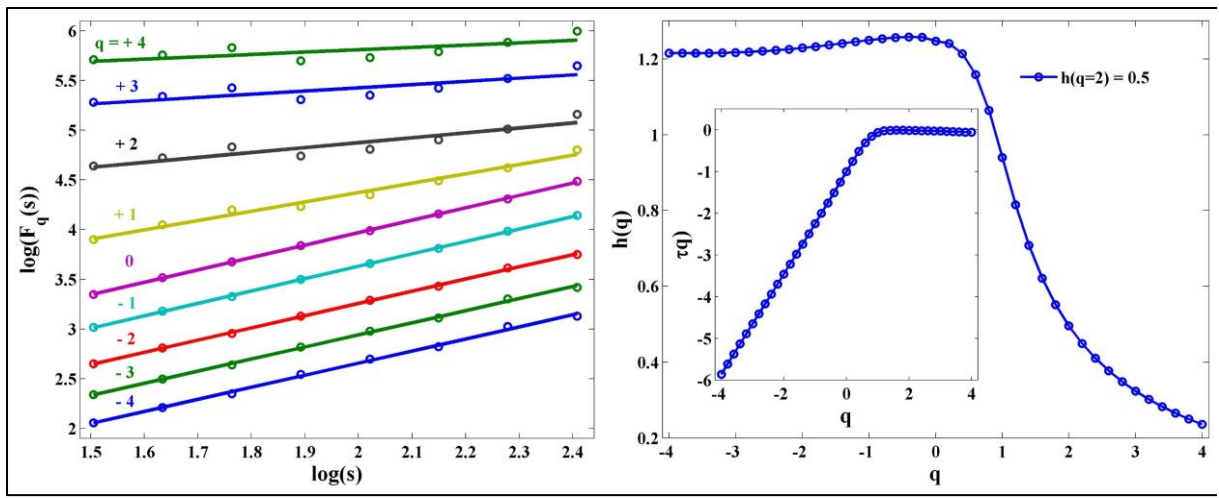

**Figure S3:** *Bi-fractal trends in the statistically equivalent behavioural random fluctuation series generated using a single power-law approximation $(P(r) \sim r^{-\alpha})$ of the Frequency vs behavioural rank data ($\alpha \sim 1.81$ corresponding to Fig. 2a).* **(a), (b):** Results of the MFDFA analysis. **(a)** The log-log (natural logarithm) plot of the moment ($q$ = -4 to +4) dependent

fluctuation function $F_q(s)$ vs length scale *s*. Stronger variations in the slopes for positive *q*-values as compared to negative *q*-values are indicative of bi-fractal scaling behaviour. **(b)** The variation of the generalized Hurst exponent $h(q)$ and the classical scaling exponent $\tau(q)$ (inset) with varying $q$. Gradual variation of $h(q)$ for $q > \alpha$ and nearly uniform values of $h(q)$ for $q \leq \alpha$ confirm the bi-fractal trends.